\documentstyle[11pt,a4]{article}

\renewcommand{\epsilon}{\varepsilon}
\begin{document}
\title{Collisional equilibrium, particle production and
the inflationary universe}
\author{ Winfried Zimdahl$^{1, 2}$,
Josep Triginer$^{1}$ and Diego Pav\'{o}n
$^{1}$\\
$^{1}$Departament de F\'{\i}sica\\
Universitat Aut\`{o}noma de Barcelona\\
08193 Bellaterra (Barcelona), Spain\\
and\\
$^{2}$Fakult\"at f\"ur Physik, Universit\"at Konstanz,
PF 5560 M678\\
D-78434  Konstanz, Germany\thanks{Present address}}
\date{\today}
\maketitle
\thispagestyle{empty}
\begin{abstract}
Particle production processes in the expanding universe
are described
within a simple kinetic model.
The equilibrium conditions for a Maxwell-Boltzmann gas
with variable
particle number are investigated.
We find that radiation and nonrelativistic matter may
be in equilibrium
at the same temperature provided the matter particles are
created at a
rate that is half the expansion rate.
Using the fact that the creation of particles is dynamically
equivalent
to a nonvanishing bulk pressure we calculate the backreaction
of this
process on the cosmological dynamics.
It turns out that the `adiabatic' creation of massive
particles with an equilibrium distribution for the
latter necessarily implies power-law inflation.
Exponential inflation in this context
is shown to become inconsistent with the second law of
thermodynamics after a time interval of the order
of the Hubble time.
\end{abstract}
\section{Introduction}
Particle production processes in the early universe are
supposed to
have influenced the cosmological history considerably.
These processes are quantum in nature \cite{BiDa}. Their
backreaction
on the
cosmological dynamics, however, are frequently studied
phenomenologically. As was observed  by Zel'dovich \cite{Zel}
and Murphy \cite{Mur} and
lateron confirmed by Hu \cite{Hu} a nonvanishing particle
production rate
is equivalent to a bulk viscous pressure in fluid cosmology.
This equivalence has been used recently in a series of
papers \cite{Prig}, \cite{Barr}, \cite{Calv}, \cite{LiGer},
\cite{ZP1}, \cite{ZPJ},
\cite{ZP2}, \cite{ZP3},  \cite{GaLeDe}, \cite{Lima},
discussing different aspects of this effective viscous
pressure approach.
Taking into account that the fluid dynamics of a gas may
be derived
from Boltzmann's equation for the one-particle distribution
function,
the problem has been addressed whether the effective viscous
pressure
approach is compatible with the kinetic theory of a gas
with varying
particle number \cite{TZP}.
A modified Boltzmann equation was proposed in which an
additional
source term describes the change of the one-particle
distribution
function due to particle number nonconserving processes,
supposedly
of quantum origin. For a simple rate approximation of
this source
term and under the assumption that the characteristic
time scale for
interactions between the particles is much smaller than
the time
scale on which the number of particles changes,
it turned out that
the effective viscous pressure approach is compatible
with kinetic
theory in homogeneous spacetimes but not in
inhomogeneous ones
\cite{TZP}.

A modification of Boltzmann's equation implies
modified conditions
for collisional equilibrium.
In the standard relativistic kinetic theory, i.e.,
without the
mentioned additional source term, it is a well
established result
\cite{Ehl},
\cite{IS}
that a simple gas can only be in `global' equilibrium
in spacetimes
that admit a timelike Killing-vector, i.e., in
stationary spacetimes.
As an immediate consequence, it is impossible to
characterize such a
gas by a `global' equilibrium distribution function
in an expanding
universe.
The only exception is a gas of massless ($m = 0$)
particles
(radiation), where the corresponding equilibrium
condition is less
restrictive and requires only the existence of a
conformal
Killing-vector.

As was shown in \cite{TZP}, the conditions for the
one-particle distribution function to have a structure
characteristic
for collisional equilibrium are weaker in the case of a
modified
Boltzmann equation, i.e., a Boltzmann equation with an
additional
source term, than in the standard case.
More specifically, for the mentioned effective rate
model of the
source term the one-particle distribution function
takes its
collisional equilibrium structure if the corresponding
spacetime
admits a conformal Killing-vector for massive particles
as well.
A gas of particles with $m > 0$ in the expanding universe
may also be
characterized by a distribution function with equilibrium
structure,
provided the number of particles is not fixed but changes
according
to a specific rate, depending on the detailed equations of
state.

The present paper provides a detailed discussion of the
conditions for collisional equilibrium of a simple gas
with nonconserved particle number.
Our main objective is to consider the possible implications
of these
solutions of the modified Boltzmann equation for the
cosmological dynamics.

Of course, the applicability of a kinetic approach
to the early stages of
the cosmological evolution is not straightforward.
A gas, however, is the only system for which the
correspondence between microscopic variables,
governed by a distribution function, and
phenomenological fluid quantities is sufficiently
well understood. All the considerations of this paper
refer to a model universe for which the kinetic approach
is assumed to be valid.
We hope that this idealized model nevertheless shares some
basic
features with our real universe.

Under these prepositions we shall establish apparently
surprizing links
between the equilibrium properties of a relativistic gas
with nonconserved particle number and the spacetime geometry
of the universe.
The most striking feature is the circumstance that the
collisional
equilibrium of a classical Boltzmann gas with varying
particle number
in some cases requires a power-law inflationary universe.

The conditions for collisional equilibrium do not fully
fix the  particle production rate.
The remaining freedom may be used to impose additional physical
requirements.
We shall discuss here two different cases and explore
the implications
of each of them for the cosmological evolution.

The first case will be that of `adiabatic' particle production,
characterized by a constant entropy per particle. The particles  
areassumed to be created with a fixed entropy. This additional
assumption
entirely determines the creation rate that is neccessary
for given
equations of state to keep the particles at equilibrium.
This rate is
the higher the more massive the particles are.
For nonrelativistic particles it is just half the expansion rate.

Now, a given creation rate is equivalent to a given effective
viscous
pressure.
With the help of the latter it is possible to calculate the
backreaction
of the production process on the cosmological dynamics.
For nonrelativistic particles
this  backreaction leads to  power-law inflation with a
behaviour of the scale factor $R$ like $R \propto t^{4/3}$ in a
homogeneous, isotropic and spatially flat universe.
Despite of the equilibrium structure of the distribution
function the generation of particles is accompanied by a
nonvanishing entropy
production.
For massive particles the comoving entropy grows as $t ^{2}$.

The second case is connected with exponential inflation.
There have been suggestions in the literature that a
sufficiently high production rate of particles or strings
might give rise to a de Sitter
phase
(\cite{Tu}, \cite{Barr}, \cite{ZP1}, \cite{ZPJ}, \cite{ZP3}).
We shall consider here the question whether this suggestion
is compatible
with the kinetic theory of a gas with variable particle number.
To this end we use the above mentioned freedom in the
production rate to
{\it impose} the condition of a constant Hubble rate and
clarify the consequences of this assumption.
We show that exponential inflation becomes
inconsistent with a
non-negative entropy production,
as required by the second law of thermodynamics,
after a time of the order of the Hubble time.

The paper is organized as follows. In section 2 we recall the
previously developed kinetic theory for particle production
and discuss the general conditions for collisional equilibrium.
Section 3 is devoted to `adiabatic' particle production and
its relation
to power-law inflation.
In section 4 we check an inflationary scenario where a
de Sitter phase
is generated by `nonadiabatic' production of gas particles
with a Maxwell-Boltzmann equilibrium distribution function
and show that it
violates the second law of thermodynamics
after about one Hubble time.
The final section 5 summarizes the main
results of the paper.\\
Units have been chosen so that $c = k_{B} =  \hbar =1$.
\section{Kinetic theory for particle production}
\subsection{General kinetic theory}
According to the discussion in \cite{TZP} we assume
that a change in the number of particles
of a relativistic gas should manifest itself in a source
term $H$ on the
level of kinetic theory.

The corresponding one-particle distribution function
$ f = f\left(x,p\right)$
of a relativistic gas with varying particle number
is supposed to obey the equation
\begin{equation}
L\left[f\right] \equiv
p^{i}f,_{i} - \Gamma^{i}_{kl}p^{k}p^{l}
\frac{\partial f}{\partial
p^{i}}
 = C\left[f\right] + H\left(x, p\right)  \mbox{ , }
\label{1}
\end{equation}
where $f\left(x, p\right) p^{k}n_{k}\mbox{d}\Sigma dP$
is the number of
particles whose world lines intersect the hypersurface element
$n_{k}d\Sigma$ around $x$, having 4-momenta in the range
$\left(p, p + \mbox{d}p\right)$;  $i$, $k$, $l$ ... = $0$,
$1$, $2$, $3$. \\
$\mbox{d}P = A(p)\delta \left(p^{i}p_{i} + m^{2}\right)
\mbox{d}P_{4}$
is the volume element on the mass shell
$p^{i}p_{i} = -  m^{2} $
in the momentum space.
$A(p) = 2$, if $p^{i}$ is future directed and
$A(p) = 0$ otherwise;
$\mbox{d}P_{4} = \sqrt{-g}\mbox{d}p^{1}
\mbox{d}p^{2}\mbox{d}p^{3}
\mbox{d}p^{4}$. \\
$C[f]$ is the Boltzmann collision term.
Its specific structure discussed e.g. by
Ehlers \cite{Ehl} will not
be relevant for our considerations.
Following Israel and Stewart \cite{IS} we shall only
require that (i) $C$
is a local function of the distribution function, i.e.,
independent
of derivatives of $f$, (ii) $C$ is consistent with
conservation of
4-momentum and number of particles, and (iii) $C$
yields a nonnegative expression for the entropy
production and does
not vanish unless $f$ has the form of a local equilibrium
distribution (see (\ref{9})).

The term $H(x,p)$ on the r.h.s. of (\ref{1})  takes into
account the fact
that the number of particles whose world lines intersect a
given
hypersurface element within a certain range of momenta may
additionally
change due to creation or decay processes, supposedly of
quantum origin.
On the level of classical kinetic theory we shall regard
this term as
a given
input quantity.
Later we shall give an example for the possible functional
structure
of $H(x, p)$.

By the splitting of the r.h.s of eq.(\ref{1}) into $C$ and
$H(x, p)$ we
have separated the collisional from the creation (decay)
events.
In this setting collisions are not accompanied by creation or
annihilation processes. In other words, once created,
the interactions between the particles
are both energy-momentum and number preserving.
For a vanishing $H(x, p)$ eq.(\ref{1})  reduces to the
familiar
Boltzmann equation (see, e.g., \cite{Ehl}, \cite{Stew}, \cite{IS},  
\cite{Groot}).

The particle number flow 4-vector
$N^{i}$ and the energy momentum tensor $T^{ik}$ are
defined in a standard way (see, e.g., \cite{Ehl}) as
\begin{equation}
N^{i} = \int \mbox{d}Pp^{i}f\left(x,p\right) \mbox{ , }
\ \ \
T^{ik} = \int \mbox{d}P p^{i}p^{k}f\left(x,p\right) \mbox{ .}
\label{2}
\end{equation}
The integrals in (\ref{2}) and throughout
the paper
are integrals over the entire mass shell
$p^{i}p_{i} = - m^{2}$.
The entropy flow vector $S^{a}$ is given by \cite{Ehl}, \cite{IS}
\begin{equation}
S^{a} = - \int p^{a}\left[
f\ln f - f\right]\mbox{d}P \mbox{ , }\label{4}
\end{equation}
where we have restricted ourselves to the case of
classical Maxwell-Boltzmann particles.

Using the general relationship \cite{Stew}
\begin{equation}
\left[\int p^{a_{1}}....p^{a_{n}}p^{b}f \mbox{d}P\right]_{;b}
= \int p^{a_{1}}...p^{a_{n}}L\left[f\right] \mbox{d}P \label{5}
\end{equation}
and eq.(\ref{1}) we find
\begin{equation}
N^{a}_{;a} = \int \left(C\left[f\right] + H\right) \mbox{d}P
\mbox{ , } \ \
T^{ak}_{\ ;k} =  \int p^{a}\left(C\left[f\right] + H\right)
\mbox{d}P
\mbox{ , }
\label{6}
\end{equation}
and
\begin{equation}
S^{a}_{;a} = - \int \ln f
\left(C\left[f\right] + H\right) \mbox{d}P
\mbox{ .} \label{8}
\end{equation}
In collisional equilibrium, which we
shall assume from
now on, $\ln f$ in
(\ref{8})
is a linear combination of the collision invariants
$1$ and $p^{a}$.
The corresponding equilibrium distribution function
becomes (see, e.g., \cite{Ehl})
\begin{equation}
f^{0}\left(x, p\right) =
\exp{\left[\alpha + \beta_{a}p^{a}\right] }
\mbox{ , }\label{9}
\end{equation}
where $\alpha = \alpha\left(x\right)$ and
$\beta_{a}$ is a timelike vector that depends on $x$ only.\\
Inserting the equilibrium function into eq.(\ref{1}) one gets
\begin{equation}
\left[p^{a}\alpha_{,a} +
\beta_{\left(a;b\right)}p^{a}p^{b}\right]f^{0}
=  H\left(x, p\right)
\mbox{ .} \label{10}
\end{equation}
\indent It is well known (\cite{Ehl}, \cite{IS})
that for $H(x, p) = 0$ this equation, which characterizes
the `global equilibrium', admits solutions
only for very special cases in which $\alpha = const$ and
$\beta_{a}$ is a timelike Killing-vector.
It will be the main objective of this paper to clarify the
consequences of
(\ref{10}) for a specific choice of the source term $H(x, p)$
to be discussed below.

With (\ref{9}), the balances (\ref{6}) reduce to
\begin{equation}
N^{a}_{;a}=\int H\mbox{d}P \mbox{\ , \ \  }
T^{ak}_{\ ;k}=\int p^{a}H\mbox{d}P \ .
\label{11}
\end{equation}

In collisional equilibrium there is entropy production only due to
the source term $H$. From eq.(\ref{8}) we obtain
\begin{equation}
S^{a}_{;a} = - \int H\left(x, p\right)
\ln f^{0} \mbox{d}P
\mbox{ , }\label{13}
\end{equation}
implying $S^{a}_{;a} = - \alpha N^{a}_{;a}
- \beta_{a}T^{ab}_{\ ;b}$.

With $f$ replaced by $f^{0}$ in
(\ref{2}) and (\ref{4}), $N^{a}$, $T^{ab}$ and $S^{a}$ may be
split with respect to the unique 4-velocity $u^{a}$ according to
\begin{equation}
N^{a} = nu^{a} \mbox{ , \ \ }
T^{ab} = \rho u^{a}u^{b} + p h^{ab} \mbox{ , \ \ }
S^{a} = nsu^{a} \mbox{  , }
\label{15}
\end{equation}
where $h ^{ab}$ is the
spatial projection tensor $h^{ab} = g^{ab} + u^{a}u^{b}$,
$n$ is the particle number density, $\rho$ is the energy
density, $p$ is the equilibrium pressure and
$s$ is the entropy per particle.
The exact integral expressions for $n$, $\rho$, $p$ and $s$ are given
by the formulae (177) - (180) in \cite{Ehl}.

Using (\ref{15}) and defining
\begin{equation}
\Gamma  \equiv \frac{1}{n} \int H\left(x, p\right) \mbox{d}P
\mbox{ , }
\label{18}
\end{equation}
the first eq.(\ref{11}) becomes
\begin{equation}
\dot{n} + \Theta n = n\Gamma  \mbox{ .}\label{19}
\end{equation}
It is obvious that $\Gamma$ is the particle production rate.
Similarly,
with the decomposition (\ref{15}) and the abbreviation
\begin{equation}
t^{a} \equiv - \int p^{a}H\left(x, p\right) \mbox{d}P \mbox{ , }
\label{20}
\end{equation}
the energy balance, following from the
second equation (\ref{11}), may be written as
\begin{equation}
\dot{\rho} + \Theta \left(\rho + p\right) - u_{a}t^{a} = 0 \mbox{ , }
\label{21}
\end{equation}
where $\Theta = u^{a}_{;a}$ is the fluid expansion.

Introducing the quantity
\begin{equation}
\pi \equiv - \frac{u_{a}t^{a}}{\Theta} \mbox{ , } \label{22}
\end{equation}
it is possible to rewrite the energy balance (\ref{21}) as
\begin{equation}
\dot{\rho} + \Theta \left(\rho + p + \pi\right) = 0 \mbox{ , }
\label{23}
\end{equation}
i.e., $\pi$ enters the energy balance in the same way as a bulk viscous
pressure does.
A rewriting like this corresponds to the introduction of an
effectively conserved energy-momentum tensor
\begin{equation}
\hat{T}^{ik}=\rho u^{i}u^{k}+(p+\pi)h^{ik} \mbox{ , }
\label{24}
\end{equation}
instead of the nonconserved quantity $T^{ik}$ of (\ref{15}).
This  interpretation of the source term (\ref{20}) as an
effective bulk
pressure was shown to be consistent for homogeneos spacetimes
\cite{TZP}.

\subsection{Effective rate approximation}
The quantity $H\left(x, p\right)$ is an
input quantity on the level of classical kinetic theory. It
supposedly represents the net effect of certain quantum
processes with variable particle numbers (see, e.g., \cite{BiDa}) at the
interface to the classical (nonquantum) level of description.
Lacking a better understanding of these processes
it was assumed in \cite{TZP}
that the influence of these processes
on the distribution function $f\left(x,
p\right)$ may be approximately
described by a linear coupling to the latter:
\begin{equation}
H\left(x, p\right) = \zeta\left(x, p\right) f^{0}\left(x, p\right)
\mbox{ .}
\label{25}
\end{equation}
Moreover,
$\zeta$ was supposed to depend on the momenta $p^{a}$ only linearly:
\begin{equation}
\zeta = - \frac{u_{a}p^{a}}{\tau\left(x\right)} +  
\nu\left(x\right)\mbox{ .}
\label{26}
\end{equation}
$\zeta$, or equivalently $\nu$ and $\tau$, characterize the rate of
change of the distribution function due to the underlying processes
with variable particle numbers.
The restriction to a linear dependence of $\zeta$ on the momentum
is equivalent to the requirement that these processes couple to the
particle number flow vector and to the energy momentum tensor
in the balances for these quantities only,
but not to higher moments of the distribution function.
This `effective rate approximation' is modeled after the relaxation time
approximations for the Boltzmann collision term (\cite{Mar},  
\cite{AW},  \cite{Maar}).
While the physical situations in both approaches are very
different, their common feature is the simplified description of
nonequilibrium phenomena by a linear equation
for the distribution function in terms of some effective
functions of space and time that characterize the relevant scales of the
process under consideration.
In the relaxation time approximation this process is determined by
the rate at which the system relaxes to an equilibrium state.
In the present case the corresponding quantity is the rate by which
the number of particles changes.

With  (\ref{25}) and (\ref{26}) for $H(x, p)$ the source terms
$n\Gamma$ and $t^{a}$ in (\ref{19}) and (\ref{21}) are given by
\begin{equation}
n\Gamma  = - \frac{u_{a}}{\tau}N^{a} +  
\nu\left(x\right)M\left(x\right)= \frac{n}{\tau } + \nu M
\mbox{ , }\label{27}
\end{equation}
where  $M$ is the zeroth moment of the distribution function,
$M \equiv \int \mbox{d}Pf\left(x, p\right)$, and
\begin{equation}
-u _{a}t^{a} = - \frac{u_{a}u_{c}T^{ca}}{\tau\left(x\right)} +  
\nu\left(x\right)u_{a}N^{a} =
- \frac{\rho}{\tau} - \nu n
\mbox{ , }\label{28}
\end{equation}
respectively.
The expression (\ref{26}) for $\zeta$ ensures that the source terms
(\ref{27}) and (\ref{28}) depend on $M$,
$N^{i}$ and $T^{ik}$ only, but not on higher moments of the
distribution function.

Using the specific structure
(\ref{25}) and (\ref{26}) of the source
term $H(x, p)$ in (\ref{10})  the
condition (\ref{10}) becomes
\begin{equation}
p^{a}\alpha_{,a}+\beta_{(a;b)}p^{a}p^{b}=\frac{E}{\tau}+\nu,
\label{29}
\end{equation}
where $E=-u_{a}p^{a}$. Decomposing $p^{a}$ according to
$p^{a}=Eu^{a}+\lambda e^{a}$, where $e^{a}$ is a unit spatial vector,
i.e., $e^{a}e_{a}=1$, $u^{a}e_{a}=0$, the mass shell condition
$p^{a}p_{a}=-m^2$ is equivalent to $\lambda^2=E^2-m^2$. The
conditions on $\alpha$ turn out to be
\begin{equation}
\dot{\alpha}=\frac{1}{\tau}\ ,\ \ \ h^{a}_{b}\alpha_{,a}=0,
\label{30}
\end{equation}
while $\beta_{a}$ obeys the equation for a conformal Killing vector
\begin{equation}
\beta_{(a;b)}=\Psi(x)g_{ab}\ , \ \ \
m^2\beta_{(a;b)}u^{a}u^{b}=\nu \ ,
\label{31}
\end{equation}
implying $\Psi = - \nu /m ^{2}$.
Different from the case $H(x, p)=0$ where $\alpha$ has to be constant in
space and time, $\alpha$ is only spatially constant in the present
case but changes along the fluid flow lines. From (\ref{31})
it follows that
$\nu=0$ for $m=0$. The condition $\nu=0$ for $m>0$, however, is
equivalent to $\beta_{(a;b)}=0$, i.e., the corresponding spacetime is
stationary.

Using $\beta_{a}=u_{a}/T$ in (\ref{31}) yields \cite{CoTu},  
\cite{Maar}, \cite{TZP}
\begin{equation}
\frac{\dot{T}}{T} = - \frac{1}{3}\Theta
\label{34}
\end{equation}
for the temperature behaviour in an expanding universe.
It is remarkable that this relation which is a consequence of the
conformal Killing-vector property for $\beta^{a}$ holds both for
$m = 0$ with $\nu = 0$ and
for $m > 0$ in the case $\nu \neq 0$.
The validity of the same equilibrium relation (\ref{34}) both for
radiation with $H(x, p) = 0$ and for massive particles with
$H(x, p) \neq 0$ has the interesting implication that
radiation and matter in the expanding Universe may be in equilibrium
at the same temperature, provided the particle number of the matter
component is allowed to change.
It is well known, that for conserved particle numbers, i.e.,
$H(x, p) = \nu
= \tau^{-1} = 0$, an equilibrium between both components is
impossible \cite{SMS,Stew}.

As a consequence of the fact that $u_{a}/T$ is a conformal,
timelike Killing-vector, the 4-acceleration $\dot{u}^{a}$ may be
expressed in terms of the spatial temperature gradient according  
to\begin{equation}
\dot{u}_{a}=-(\ln T)_{,b}h^{b}_{a}.
\label{35}
\end{equation}
Within the linear theory of irreversible processes
this coincides with the condition for the
heat flow to vanish (see, e.g. \cite{Steph}).
For a comoving observer in the rotation free case (\ref{35})  
reduces to\cite{TZP}
\begin{equation}
(T \sqrt{-g_{00}})_{,\nu}=0 \mbox{ , } (\nu = 1,2,3)
\mbox{ .}\label{36}
\end{equation}
The latter formula which states that the quantity
$T \sqrt{-g_{00}}$
has to be spatially constant if particles
with $m > 0$ are produced,
replaces Tolman's relation
$(T \sqrt{-g_{00}})_{,n}=0$, according to which
$T \sqrt{-g_{00}}$ is constant both in space and time
in case $u_{a}/T$ is a Killing-vector.

The conformal Killing-equation (\ref{31}) provided us with the
temperature law (\ref{34}).
On the other hand, the behaviour of the temperature is determined
thermodynamically.
The fluid equations of state may generally be written as
\begin{equation}
p = p\left(n, T\right) \mbox{ , \ \ } \rho = \rho\left(n, T\right)
\mbox{ .}\label{37}
\end{equation}
Differentiating the latter relation and using the
balances (\ref{19}) and (\ref{23})
one finds
\begin{equation}
\frac{\dot{T}}{T} = - \Theta
\left[\frac{\partial p/\partial T}{\partial \rho/\partial T}
+ \frac{\pi}{T\partial \rho/\partial T}\right]
+ \Gamma
\left[\frac{\partial p/\partial T}{\partial \rho/\partial T}
- \frac{\rho + p}{T\partial \rho/\partial T}\right]
\mbox{ ,}\label{39}
\end{equation}
where $\partial p/\partial T \equiv
\left(\partial p/\partial T\right)_{n}$ and
$\partial \rho /\partial T \equiv
\left(\partial \rho /\partial T\right)_{n}$.
Relation (\ref{39}) was first derived in \cite{Calv}.

In the specific case of  a classical gas
the equations of state are
$ p =  n T$ with
\cite{Groot}
\begin{equation}
n =  \frac{4\pi m^{2}T}
{\left(2\pi\right)^{3}}K_{2}\left(
\frac{m}{T}\right) \exp{\left[\alpha\right]}
\mbox{ , }\label{41}
\end{equation}
and $\rho =  n e(T)$ with
\begin{equation}
e =   m\frac{K_{1}\left(\frac{m}{T}\right)}
{K_{2}\left(\frac{m}{T}\right)} + 3 T =
m\frac{K_{3}\left(\frac{m}{T}\right)}
{K_{2}\left(\frac{m}{T}\right)} -   T
\mbox{ .}\label{42}
\end{equation}
The quantities $K_{n}$ are modified Bessel-functions
of the second
kind \cite{Groot}.
Employing the well-known differential and
recurrence relations for the latter,
one gets
\begin{equation}
\frac{\partial \rho }{\partial p} \equiv
\frac{\partial \rho/\partial T}{\partial p/\partial T} =
z^{2} - 1 + 5\frac{h}{T}
- \left(\frac{h}{T}\right)^{2}
\mbox{ , }\label{46}
\end{equation}
where $z = m/T$ and
\begin{equation}
h = e + \frac{p}{n} = m \frac{K_{3}\left(z\right)}
{K_{2}\left(z\right)}
\mbox{ , }\label{47}
\end{equation}
is the enthalpy per particle.

By virtue of (\ref{22}), (\ref{27}) and (\ref{28}), the temperature  
law (\ref{39}) may be expressed in terms of $\nu$ and $\tau$.  
Because of
$\partial p/\partial T = p/T$ the $\tau$-terms cancel and we obtain
\begin{equation}
\frac{\dot{T}}{T} = - \Theta
\frac{\partial p/\partial T}{\partial \rho/\partial T}
 +
\frac{\nu n}{T \partial \rho /\partial T}
\left[1 - \frac{\rho }{n}\frac{M}{n}\right]
\mbox{ .}\label{48}
\end{equation}
Combining the latter expression with (\ref{34}) yields
\begin{equation}
\nu = T \frac{1 - \frac{1}{3}\frac{\partial \rho}{\partial p}}
{1 - \frac{\rho }{n}\frac{M}{n}} \Theta
\mbox{ .}\label{49}
\end{equation}
The $\nu$-part of the source terms (\ref{26}), (\ref{27}) and
(\ref{28})
is fixed by the thermodynamic functions of the gas.
The zeroth moment $M$ of the classical distribution function
$f^{0}$ is given by
\begin{equation}
M = \frac{4 \pi m T}{\left(2 \pi   \right) ^{3}}
K _{1} \left(\frac{m}{T}   \right) \exp \left[  \alpha \right]
\mbox{ .}\label{50}
\end{equation}

There exists an alternative formula for the function $\nu$ that
may be obtained from the Gibbs-Duhem equation
\begin{equation}
\mbox{d} p = n s \mbox{d}T +  n \mbox{d}\mu  \mbox{ .}
\label{a}
\end{equation}
With \cite{Groot}
\begin{equation}
n s = \frac{\rho + p}{T} - \frac{n \mu}{T}
\label{b}
\end{equation}
one finds
\begin{equation}
\dot{p} = \left(\rho + p\right)\frac{\dot{T}}{T}
+ n T \left(\frac{\mu }{T}\right)^{\displaystyle \cdot} \mbox{ .}
\label{c}
\end{equation}
Using here $p = n T$ yields
\begin{equation}
\left(\frac{\mu }{T} \right)^{\displaystyle \cdot}
= \frac{\dot{n}}{n} - \frac{\rho }{p}\frac{\dot{T}}{T} \mbox{ .}
\label{d}
\end{equation}
Applying (\ref{19}) with (\ref{27}) for $\dot{n}/n$, eq.(\ref{34})  
for$\dot{T}/T$ and the first of eqs.(\ref{30}) for
$\dot{\alpha} \equiv \left(\mu /T\right) ^{\displaystyle \cdot}$,  
wearrive at
\begin{equation}
\nu = \frac{n}{M} \left(1 - \frac{\rho}{3p} \right) \Theta  =
\frac{m}{3} \frac{K _{2}}{K _{1}}
\left[4 - z \frac{K _{3}}{K _{2}}\right]
\Theta \mbox{ .}  \label{51}
\end{equation}
The equivalence between the expressions
(\ref{49}) and (\ref{51})
becomes manifest if one uses the recurrence
relations for the functions $K _{n}$
(see, e.g., \cite{Groot})
to replace, e.g.,
$K _{1}\left(z\right)$ in $M$ in (\ref{49}) by $K _{2}$ and $K _{3}$. 
With $\nu$ given either by (\ref{49}) or (\ref{51}),
the conformal factor $ \Psi $ in (\ref{31}) is fixed
as well:
\begin{equation}
\Psi  = -
\frac{1}{3m} \frac{K _{2}}{K _{1}}
\left[4 - z \frac{K _{3}}{K _{2}}\right]
\Theta \mbox{ , } \label{52}
\end{equation}
i.e., it is completely determined by the thermodynamic
functions of the gas.\\ In the limit $z \gg 1$ we have
$\nu \approx - m z \Theta /3 $
and a corresponding expression for the conformal factor.
While $\nu$ is uniquely  determined by the condition (\ref{34}),
the function $\tau$ is still arbitrary. We are free to impose
a further condition to fix $\tau$.
\section{Adiabatic particle production and power-law inflation}
\subsection{The creation rate}
In \cite{TZP} as well as in earlier papers (\cite{Calv},  
\cite{LiGer}, \cite{ZP1}, \cite{ZPJ}, \cite{ZP2}, \cite{ZP3}) the  
particle
production was assumed to be `adiabatic'. This implies that all
particles are amenable to a perfect fluid description immediately
after their creation.
The particles are produced with a fixed entropy, i.e., the entropy
per particle $s$ does not change. The corresponding condition
$\dot{s} = 0$ determines $\tau$.

From the Gibbs equation
\begin{equation}
T\mbox{d}s = \mbox{d}\frac{\rho}{n} + p\mbox{d}\frac{1}{n}
\label{54}
\end{equation}
together with (\ref{19}) and (\ref{21}) we find
\begin{equation}
nT\dot{s} = u_{a}t^{a} - \left(\rho + p\right)\Gamma \mbox{ .}
\label{55}
\end{equation}
According to (\ref{55})
the condition $\dot{s} = 0$ is generally equivalent to
\begin{equation}
u_{a}t^{a} = \left(\rho + p\right)\Gamma \mbox{ , }
\label{56}
\end{equation}
relating the source term in the energy balance to that in the
particle number balance.
Using (\ref{27}) and (\ref{28}) with (\ref{49}) in
(\ref{56}) provides us with
\begin{equation}
\tau ^{-1} = \left(1 - \frac{1}{3}\frac{\partial \rho }
{\partial p}\right)
\frac{1 - \frac{\rho  + p}{n}\frac{M}{n}}
{1 - \frac{\rho }{n}\frac{M}{n}} \Theta
\mbox{ .}\label{57}
\end{equation}
Because of (\ref{27}) the particle production rate
$\Gamma $ turns out to be
\begin{equation}
\Gamma  = \left[1 - \frac{1}{3}\frac{\partial \rho }
{\partial p} \right] \Theta
\label{58}
\end{equation}
in this case, equivalent to
\begin{equation}
\pi   = - \left[1 - \frac{1}{3}\frac{\partial \rho }
{\partial p} \right]
\left(\rho + p\right)
\mbox{ , }\label{59}
\end{equation}
where we have used (\ref{22}).
In fact, expression (\ref{58}) for the
adiabatic particle production rate may be obtained
from (\ref{34}), (\ref{39}), (\ref{56}) and (\ref{22})
without explicitly knowing $\nu$ and $\tau $.
A separate knowledge of $\nu$ and $\tau $, however,
will be needed in section 4 below.

Inserting into (\ref{58}) the expression (\ref{46})
for $\partial \rho/\partial p$  we have
\begin{equation}
\Gamma =  \left[1 - \frac{1}{3}
\left(z^{2} - 1 + 5\frac{h}{T}
- \left(\frac{h}{T}\right)^{2}\right)
\right]
\Theta
\mbox{ , }
\label{60}
\end{equation}
or an equivalent expression for $\pi$.
In the limiting case $m = 0$ (radiation) with $h = 4 T$, the
relation (\ref{60}) yields $\Gamma = 0$ in agreement with the
result of \cite{TZP} that the adiabatic production of massless
particles
is forbidden if $\beta^{a}$ is a conformal Killing-vector.
In the opposite limiting case $z \gg 1$ the enthalpy per particle  
is\begin{equation}
h \approx  \left[z  + \frac{5}{2}
+ \frac{15}{8}z^{-1}\right] T
\mbox{ , }\label{61}
\end{equation}
yielding
\begin{equation}
\Gamma \approx  \frac{1}{2}\Theta
\mbox{ , }\label{62}
\end{equation}
i.e., the production rate for massive particles in equilibrium is  
half the expansion rate.
Equivalently, since the Hubble factor of the Robertson-Walker  
metric is $H \equiv \Theta/3$,
the characteristic time-scale $\Gamma ^{-1}$ for the production of
very massive
particles, neccessary for an equilibrium distribution of the latter,
is $2/3$ of the Hubble time $H ^{-1}$ in the case $\dot{s} =0$.
\subsection{Backreaction on the cosmological dynamics}
The main advantage of a fluid approach to particle production is  
the possibility to calculate the backreaction of this process
on the cosmological dynamics.
For homogeneous spacetimes this problem is equivalent to studying the
dynamics of a bulk viscous fluid universe \cite{TZP}.
While generally the determination of the bulk pressure is an involved
problem on its own we are in a much better situation
in the present case.
The effective bulk pressure here is completely fixed by the conformal
Killing-vector conditions (\ref{31}), the adiabaticity
condition $\dot{s} = 0$ and the equations
of state.
Using (\ref{59}) in the energy balance (\ref{23}), the latter takes the
form
\begin{equation}
\dot{\rho} = - \Theta \frac{\rho + p}{3}
\frac{\partial \rho}{\partial p}
\mbox{ , }\label{63}
\end{equation}
with $\partial \rho/\partial p$ given by (\ref{46}).
While for $m = 0$, equivalent to $p = \rho/3$, with $\pi = 0$ we
recover the familiar behaviour for radiation, the opposite limiting
case, $z \gg 1$, yields
\begin{equation}
\dot{\rho} = - \frac{3}{2} H \rho
\mbox{ , }\label{64}
\end{equation}
or, with $H = \dot{R}/R$, where $R$ is the scale factor of the
Robertson-Walker metric,
\begin{equation}
\rho \propto R^{-3/2}
\mbox{ .}\label{65}
\end{equation}
The energy density decreases less than without particle production
($\rho \propto R^{-3}$).
A parallel statement  holds for $n$. From (\ref{19}) with  
(\ref{62}) we find
$n \propto R^{-3/2}$
instead of $n \propto R^{-3}$ for $\Gamma = 0$.

We recall that the temperature behaviour is given by (\ref{34}),
irrespective of the equations of state.

Restricting ourselves to a homogeneous, isotropic and spatially flat
universe with
\begin{equation}
3 \frac{\dot{R}^{2}}{R^{2}} = \kappa \rho
\mbox{ , }\label{66}
\end{equation}
where $\kappa$ is Einstein's gravitational constant, and
\begin{equation}
\frac{\dot{H}}{H} = \frac{1}{2}\frac{\dot{\rho}}{\rho}
\mbox{ , }\label{67}
\end{equation}
where the latter follows by virtue of
$3 \ddot{R}/R = - \left(\kappa /2\right)
\left(\rho + 3p + 3 \pi \right)$,
we find with the help of
(\ref{64}) that the scale factor $R$ behaves like
\begin{equation}
R \propto t^{4/3}
\mbox{ , }\label{68}
\end{equation}
instead of the familiar $R \propto t^{2/3}$ for
$\rho \propto R^{-3}$, i.e., for $\Gamma = 0$.

If massive particles are dynamically dominating, the adiabatic
production rate
for these particles that is neccessary to keep them governed by an
equilibrium distribution function backreacts on the dynamics of the
Universe in a way that implies power-law inflation.
In other words, {\it massive particles with nonconserved particle
number and fixed entropy per particle
are allowed to be in collisional equilibrium only in a power-law
inflationary universe}.

The backreaction is largest for $z \gg 1$ and it vanishes for $m =
0$. In the latter case the familiar $R \propto t^{1/2}$ behaviour for
radiation is recovered.
There exists an intermediate equation of state yielding $R \propto t$
with $\ddot{R} = 0$.
The obvious condition for $\ddot{R} = 0$ is
\begin{equation}
\rho + 3p + 3\pi = 0
\mbox{ .}\label{70}
\end{equation}
Using the expression (\ref{59}) for $\pi$, this condition reduces
to
\begin{equation}
\frac{\partial \rho}{\partial p} = \frac{2 \rho }{\rho + p}
\mbox{ .}\label{71}
\end{equation}
Inserting here (\ref{46}), we obtain the following cubic  equation
for $h/T$:
\begin{equation}
\left(\frac{h}{T}\right)^{3}
-  5 \left(\frac{h}{T}\right)^{2} - \left(z ^{2} - 3\right)
\frac{h}{T} - 2 = 0
\mbox{ .}\label{72}
\end{equation}
Together with (\ref{47}) the latter equation determines the  
criticalvalue   $z_{cr}$ that characterizes the case $\ddot{R} = 0$.
The numerical result is $z _{cr} \approx 9.55 $.
We have $\ddot{R} > 0$ for $z > z_{cr}$
while  $z < z_{cr}$ corresponds
to $\ddot{R} < 0$.

In terms of a `$\gamma$'-law, i.e., $p = (\gamma - 1)\rho$, the
corresponding
critical $\gamma$ value $\gamma _{cr}$ is
\begin{equation}
\gamma_{cr} = \frac{z_{cr}\frac{K_{3}\left(z_{cr}\right)}
{K_{2}\left(z_{cr}\right)}}
{z_{cr}\frac{K_{3}\left(z_{cr}\right)}
{K_{2}\left(z_{cr}\right)} - 1}  \approx 1.09
\mbox{ .}\label{77}
\end{equation}
There is power-law inflation for $1 \leq \gamma < \gamma_{cr}$.
The case $\gamma = \gamma_{cr}$ corresponds to $\ddot{R} = 0$.
\subsection{Entropy production}
It is a specific feature of our approach that a gas, kept in
equilibrium in the expanding universe by a nonvanishing particle
creation rate exhibits
a nonvanishing entropy production density
\begin{equation}
S ^{a}_{;a} = n \Gamma s
\mbox{ , }\label{78}
\end{equation}
with $\Gamma $ from (\ref{58}).
While there is no entropy production due
to conventional dissipative
processes, it is the particle production rate itself
that is connected with an increase in the entropy. \\
The comoving entropy $\Sigma $ is defined by
$\Sigma = n s R^{3}$.
This quantity may vary either by a change in the entropy per
particle $s$, or by a varying number $N \equiv nR^{3}$ of  
particles.The first possibility was excluded here by the requirement  
(\ref{56}).The second one, an increase in the number of $m > 0$  
particles,
however, was necessary for the particles to be governed by an
equilibrium distribution function.
Consequently, this kind of equilibrium even implies an increase in the
comoving entropy.

While the latter statement sounds unfamiliar and does
never hold indeed for systems with conserved particle numbers, it is
a natural outcome for a nonvanishing creation rate $\Gamma$.

A further unfamiliarity, related to the previous one, is that
one has nonzero entropy production in the
corresponding power-law inflationary phase.
In the standard inflationary scenarios  the phases of accelerated
expansion are not accompanied by an increase in $\Sigma$ and all
the entropy is produced during a
subsequent reheating period.
In our approach one finds from (\ref{19}), (\ref{62}) and  
(\ref{68})in  the limiting case of very massive particles $z \gg 1$,
that  $\Sigma \sim t^{2}$, i.e., the
comoving entropy grows quadratically with the cosmic time.
\section{Exponential inflation}
The functions $\nu$ and $\tau$ characterizing the particle  
productionprocess were considered to be input quantities on the  
level of classical kinetic theory.
Only the specific expression (\ref{49}) for $\nu$, however,  
guarantees the fulfilment of the equilibrium conditions (\ref{30}) -  
(\ref{31}).
Once the remaining freedom in $\tau$ is fixed by some physical  
requirement
like that of the `adiabaticity' of the creation process in the  
preceeding section, the dynamics of our model universe is uniquely  
determined.
While the `adiabaticity' condition appears `natural' it is not the  
only possible choice.
In the present section we use the freedom in $\tau$ to {\it impose}  
the condition of a constant Hubble rate, i.e., the condition for  
exponential inflation, and check whether this is consistent with   
the
equilibrium solution of the (modified) Boltzmann equation.
As was already mentioned in the introduction, the possibility of
a particle- or string-production-driven
de Sitter phase has recently attracted
some interest (\cite{Tu}, \cite{Barr}, \cite{ZP1}, \cite{ZPJ},  
\cite{ZP3}).
We shall show here that this feature is not consistent with the
kinetic theory of a Maxwell-Boltzmann gas in collisional equilibrium.

It is obvious from (\ref{23}) and (\ref{67}) that the condition
$H = H _{0} = const$ for exponential inflation is
\begin{equation}
\pi = - \left(\rho + p \right)
\mbox{ .}\label{80}
\end{equation}
On the other hand, by (\ref{22}) and (\ref{28}),
the effective viscous pressure is given in terms of $\nu$ and  
$\tau$.Using either (\ref{51}) or (\ref{49})
for $\nu$ and combining (\ref{22}) and (\ref{28}) fixes $\tau$ to be
\begin{equation}
\tau ^{-1} = \left(\gamma + \frac{z}{3}\frac{nm}{\rho}\right)  
\Theta= \left[\gamma - \frac{p}{\rho }
\frac{1 - \frac{1}{3} \frac{\partial \rho }{\partial p}}
{1 - \frac{\rho }{n}\frac{M}{n}}\right] \Theta
\mbox{ .}\label{81}
\end{equation}
Solving (\ref{55}) with (\ref{27}) and (\ref{28}) for
$\tau ^{-1}$ provides us with
\begin{equation}
\tau ^{-1} = \left(1 - \frac{1}{3}\frac{\partial \rho}
{\partial p} \right)
\frac{1 - \frac{\rho + p}{n}\frac{M}{n}}{1 -
\frac{\rho}{n}\frac{M}{n}} \Theta
 - \dot{s}
\mbox{ , }\label{82}
\end{equation}
which differs from (\ref{57}) by the nonvanishing
$\dot{s}$-term.
Eliminating $\tau ^{-1}$ by (\ref{81}) yields
\begin{equation}
\dot{s} = - \frac{\gamma}{3}\frac{\partial \rho}{\partial p}
\Theta
\label{83}
\end{equation}
for the change of the entropy per particle.
The condition for exponential inflation implies a
negative $\dot{s}$ !
As we shall show below, this decrease in the entropy
per particle is compatible
with the second law of thermodynamics only up to time
scales of the order of the Hubble time.
For larger times
exponential inflation will turn out to be forbidden
thermodynamically.

Instead of (\ref{78}) we now have
\begin{equation}
S^{a}_{;a} = n s \Gamma  + n \dot{s}
\label{84}
\end{equation}
for the entropy production rate.
The particle production rate $\Gamma $ itself depends
on $\dot{s}$ according to
\begin{equation}
\Gamma  = \left(1 -
\frac{1}{3}
\frac{\partial \rho}{\partial p}\right) \Theta - \dot{s}
\mbox{ , }\label{85}
\end{equation}
as follows from (\ref{27}) with (\ref{49}) and (\ref{82}).
A negative
$\dot{s} $ enlarges the production rate $\Gamma $
compared with the case
$\dot{s} = 0 $.
Inserting (\ref{83}) into (\ref{85}) it follows
\begin{equation}
\Gamma  = \left(1 + \frac{1}{3}\frac{\partial \rho}
{\partial p}
\frac{p}{\rho}\right) \Theta
\mbox{ .}\label{86}
\end{equation}
The entropy production density becomes
\begin{equation}
S ^{a}_{;a} = n s \Theta  - \frac{n \mu }{T}
\frac{p}{3 \rho }
\frac{\partial \rho }{\partial p}\Theta
\mbox{ , }\label{87}
\end{equation}
where the expression (\ref{b}) has been used.
$\mu$ is the chemical potential with $\mu /T \equiv \alpha$.

Equations (\ref{19}) and (\ref{86}) imply
\begin{equation}
\frac{\dot{n}}{n} = \frac{1}{3}\frac{\partial \rho}
{\partial p}\frac{p}{\rho}
\Theta
\mbox{ , }\label{89}
\end{equation}
i.e.,
the number density
increases during the period of exponential inflation.
This is a natural consequence of the fact that the particle
production rate (\ref{86})
is larger than the expansion rate.

We point out that the considerations of this section are not
restricted to massive particles.
Different from the adiabatic case of the previous section
that did not allow for the production of ultrarelativistic
particles with
$z \rightarrow 0$, there is no corresponding restriction
in an exponentially inflating universe.
The reason for this is a simple one: Since (\ref{48}) does
not depend on $\tau$, any choice of the latter may be
compatible with the equilibrium conditions.
It was the adiabaticity condition of the previous section that
introduced a proportionality between $\tau ^{-1}$ and $\nu$.
Only by virtue of this additional requirement $\tau$ was
related to $\nu$.
Since $\nu$ neccessarily vanishes for $z \rightarrow 0$ the
condition for adiabatic particle production implies
$\tau ^{-1} \rightarrow 0$ as well.
In the present case of a constant Hubble rate the corresponding
condition on $\tau$ is independent of $\nu$. Consequently,
even for vanishing $\nu$ a nonzero particle production
due to $\tau ^{-1} > 0$, that does not affect the
behaviour (\ref{48}) of the temperature, is possible.

With $p _{r} = \rho _{r}/3$, where the subscript `$r$'
refers to radiation,  equation (\ref{89}) yields
\begin{equation}
\frac{\dot{n}_{r}}{n _{r}} = H _{0}
\mbox{ , }\label{90}
\end{equation}
equivalent to $n _{r} \propto R$. Since according to
(\ref{34}) we have $T _{r} \propto R ^{-1}$ always,
this behaviour is consistent with
$\rho _{r} = 3n _{r}T _{r} = const$ in the de Sitter phase.

With $\nu _{r} = 0$ for radiation, the entropy per
particle changes according
to (\ref{55}) with (\ref{27}) and (\ref{28}) as
\begin{equation}
\dot{s}_{r} = - \frac{1}{\tau _{r}}
\mbox{ .}\label{91}
\end{equation}
Taking into account (\ref{b})
and $\rho _{r} + p _{r} = 4 n _{r} T _{r}$, the
expression (\ref{91}) is compatible  with $\alpha _{r}
\equiv \mu _{r}/T _{r}$ and the first equation (\ref{30}).
Furthermore, in the ultrarelativistic case the relations
\begin{equation}
\Gamma _{r} = \frac{1}{\tau _{r}}
\mbox{ , }\label{92}
\end{equation}
and
\begin{equation}
S ^{a}_{;a} = \frac{n _{r}}{\tau _{r}}\left(s _{r} - 1\right)
\mbox{ , }\label{93}
\end{equation}
are valid.
With $\nu _{r} = 0$ for $p _{r} = \rho _{r}/3$ and
using (\ref{22}), (\ref{55}) and (\ref{27}),
the condition (\ref{80}) for exponential inflation
reduces to
\begin{equation}
\tau _{r} = \frac{1}{4}H ^{-1}_{0}
\mbox{ .}\label{94}
\end{equation}
The characteristic time scale for particle
production is a quarter of the Hubble time.
Combining (\ref{b}) and (\ref{94}), the entropy
production density (\ref{93}) may be written as
\begin{equation}
S ^{a}_{;a} = 4 n _{r} H _{0}\left(3 -
\frac{\mu _{r}}{T _{r}}\right)
\mbox{ .}\label{95}
\end{equation}
The quantity $\mu _{r}/ T _{r}$ is given by \cite{Groot}
\begin{equation}
\frac{\mu _{r}}{T _{r}} =
\ln \left[\frac{\pi ^{2}}{T _{r} ^{3}}n _{r}\right]
\mbox{ .}\label{96}
\end{equation}
While $\mu _{r}/T _{r}$ is constant for $\Gamma = 0$,
this is no longer true in the present case.
From (\ref{b}), (\ref{91}) and (\ref{94}) we have
\begin{equation}
\left(\frac{\mu _{r}}{T _{r}}\right)^{\displaystyle \cdot}
= 4 H _{0}
\mbox{ , }\label{97}
\end{equation}
and, after integration,
\begin{equation}
\frac{\mu _{r}}{T _{r}} =  \left(\frac{\mu _{r}}{T  
_{r}}\right)_{0}+ 4 H _{0}\left(t - t _{0}\right)
\mbox{ .}\label{98}
\end{equation}
The first term on the r.h.s. corresponds to the
initial value at $t = t _{0}$.
The ratio $\mu _{r}/T _{r}$ grows linearly with $t$.
From (\ref{98}) and (\ref{95}) it is obvious that
the fulfillment of the condition
$S ^{a}_{;a} \geq 0$ is only guaranteed for
\begin{equation}
t - t _{0} \leq \frac{3}{4} H _{0}^{-1}
\left[1 - \frac{1}{3} \left(\frac{\mu _{r}}
{T _{r}}\right)_{0} \right]
\mbox{ .}
\label{99}
\end{equation}
After a time of the order of the expansion time
the second law becomes violated, i.e.,
the corresponding process is forbidden.

Denoting the opposite limiting case
$z \gg 1$ for nonrelativistic matter by the subscript `$m$',
we find from (\ref{89}) the expression
\begin{equation}
\frac{\dot{n}_{m}}{n _{m}} \approx
\frac{1}{2}\frac{T _{m}}{m}\Theta
\mbox{ , }
\label{100}
\end{equation}
for the time dependence of the particle number density,
i.e., in the limit $z \gg 1$ the number density is almost constant. 
The change in the entropy per particle is
\begin{equation}
\dot{s}_{m} = - \frac{\Theta}{2}
\mbox{ .}\label{101}
\end{equation}
The quantities $\tau _{m}^{-1}$ and $\nu _{m}$ are given by
\begin{equation}
\tau _{m}^{-1} \approx \left(\frac{z}{3} +  
\frac{1}{2}\right)\Theta\label{102}
\end{equation}
and
\begin{equation}
\nu _{m}\frac{M}{n _{m}} \approx -
\left(\frac{z}{3} - \frac{1}{2}\right)\Theta
\mbox{ , }\label{103}
\end{equation}
respectively.
Although $\tau _{m}^{-1} \gg \Theta$ and
$\mid \nu \mid M/n _{m}\gg \Theta $, the
terms proportional to $z$ cancel in the
expression (\ref{27}) for $\Gamma $, yielding
\begin{equation}
\Gamma _{m} \approx \Theta
\mbox{ , }\label{104}
\end{equation}
in agreement with the corresponding result from (\ref{86}).
Introducing $\rho _{m} = n _{m} m + 3 n _{m} T _{m}/2$ and
$p _{m} = n _{m}T _{m}$ into (\ref{b}),
the expressions (\ref{101}) and (\ref{102})
are  consistent with
$\left(\mu _{m}/T _{m}\right)^{\displaystyle
\cdot} \equiv \dot{\alpha}_{m} = \tau _{m}^{-1}$,
as required by (\ref{30}).
The growth rate of $\alpha _{m}$ is much
larger than the particle production rate (\ref{104}).

The entropy production density in the case $z \gg 1$ becomes
\begin{equation}
S ^{a}_{;a} \approx \left[\left(z - \frac{\mu _{m}}{T _{m}}\right)
+ \frac{5}{2}\right]n _{m} \Theta
\mbox{ .}\label{105}
\end{equation}

The chemical potential, the number density and $z$
are related by \cite{Groot}
\begin{equation}
\exp \left[\mu/T \right] =
\frac{2 \pi ^{2}}{m ^{2}TK _{2}\left(z\right)} n
\mbox{ .}\label{106}
\end{equation}
Using for $K _{2}\left(z\right)$
the asymptotic representation for $z \gg 1$
\begin{equation}
K _{2}\left(z\right) \approx
\sqrt{\frac{\pi }{2 z}} \exp \left[-z\right]
\mbox{ , }\label{107}
\end{equation}
one finds
\begin{equation}
z - \frac{\mu _{m}}{T _{m}} = -
\ln \left[\left( \frac{2 \pi }{m T _{m}}\right)
^{3/2} n _{m}\right]
\mbox{ .}\label{108}
\end{equation}
For $\Gamma = 0$ one has $n _{m} \propto R ^{-3}$
and $T _{m} \propto R ^{-2}$ and
the quantity $z - \mu _{m}/ T _{m}$ is a constant.
In the present case we find from (\ref{b}) and (\ref{101})
\begin{equation}
\dot{s} = \left(z -\frac{\mu _{m}}
{T _{m}} \right)^{\displaystyle \cdot}
= - \frac{3}{2} H _{0}
\mbox{ .}\label{109}
\end{equation}
Integration yields
\begin{equation}
z - \frac{\mu _{m}}{T _{m}} =
\left(z -\frac{\mu _{m}}{T _{m}} \right)_{0}
- \frac{3}{2} H _{0} \left(t - t _{0}\right)
\mbox{ .}\label{110}
\end{equation}
After a time interval
\begin{equation}
t - t _{0} \approx \frac{5}{3} H _{0}^{-1}
\left[1 + \frac{2}{5} \left(z -
\frac{\mu _{m}}{T _{m}} \right)_{0}\right]
\label{111}
\end{equation}
the condition $S ^{a}_{;a} \geq 0$ becomes violated,
i.e., the production of massive particles is not
consistent either with exponential inflation.

These results show explicitly, that the production of
Maxwell-Boltzmann particles in equilibrium is
compatible with a de Sitter phase
only on time scales of the order of the Hubble time.

In order to avoid misunderstandings we point out
that this statement does not necessarily imply
that all the above mentioned fluid cosmological scenarios
(\cite{Tu}, \cite{Barr}, \cite{ZP1}, \cite{ZPJ},
\cite{ZP3}) are wrong.
We have only excluded the case that the condition
(\ref{80}) for exponential inflation and the
equilibrium condition (\ref{34}) are fulfilled simultaneously.
If the equilibrium conditions (\ref{30}) - (\ref{31})
are dropped, the conditions for exponential inflation
are less restrictive.
Retaining the equilibrium conditions corresponds to
the so-called `global' equilibrium, while dropping
them and having nevertheless an equilibrium structure
of the distribution function
corresponds to the `local' equilibrium case
(\cite{Maar}, \cite{TZP}).
Taking into account deviations from either of the
equilibrium distribution
functions implies an imperfect fluid universe which
requires separate investigations (\cite{RoyM}, \cite{Zim}).

Instabilities of the de Sitter universe  due
to quantum processes (\cite{Star,Ford}) or due to
viscous pressures, modelling particle production
phenomena (\cite{JBar}) are well known features.
A breakdown of the de Sitter phase due to a
violation of the condition   $S ^{a}_{;a} \geq 0$
as in the present paper may be regarded as
thermodynamical instability.
If the classical evolution of our model universe
is assumed to start in a `global' equilibrium
de Sitter phase, our results imply the latter to
become thermodynamically unstable after a time
of the order of the Hubble time.
On the basis of our setting there exist two
qualitative
possibilities for a subsequent evolution of
the universe in accordance with the second law
of thermodynamics.
According to the first possibility the universe
may leave the `global' equilibrium state and  stay
in a `local' or nonequilibrium de Sitter phase
(\cite{RoyM,Zim}).
Alternatively, the universe may leave the
de Sitter phase altogether but remain in
`global' equilibrium, being characterized, e.g.,
by adiabatic particle production, including
the possibility of power-law inflation as in
section 3  above.
Whether one of these scenarios
will actually be realized, can only
be decided once reliable knowledge about the functions
$\nu$ and $\tau $, that characterize the
particle production, is available from the
underlying quantum physics.
\section{Conclusions}
Within an `effective rate model' we have studied the
equilibrium conditions for a Maxwell-Boltzmann gas
with variable particle number.
Our main findings can be summarized as:
1. Radiation  and nonrelativistic matter may be in
collisional equilibrium at the same temperature
provided, the number of matter particles increases
at a specific rate.
2. Collisional equilibrium for massive particles
with adiabatically  increasing particle number is
only possible in a power-law inflationary universe.
3. Exponential inflation is consistent with the
`global'
equilibrium conditions for a Maxwell-Boltzmann gas
only for a time interval of the order of the Huble time.
For larger times a `global' equilibrium de Sitter
phase violates the second law of thermodynamics.    \\
\ \\
{\bf Acknowledgement}\\
This paper was supported by the
Deutsche Forschungsgemeinschaft,
the Spanish Ministry of Education
(grant PB94-0718) and the NATO
(grant CRG 940598).
J.T. acknowledges support of the FPI grant
AP92-39172486.

\ \\

\end{document}